\documentclass[twocolumn,aps,pre,preprintnumbers,amsmath,amssymb,nofootinbib]{revtex4-1}

\usepackage{graphicx}
\usepackage{dcolumn}
\usepackage{bm}
\usepackage{textcomp}
\usepackage{booktabs}
\usepackage{color}

\begin{document}

\title{Coarse-grained molecular dynamics simulation \\ of binary charged lipid membranes: \\ 
Phase separation and morphological dynamics}

\author{Hiroaki Ito}
\email{ito@hh.mech.eng.osaka-u.ac.jp}
\affiliation{%
Department of Mechanical Engineering, Graduate School of Engineering, Osaka University, Osaka 565-0871, Japan
}%
\author{Yuji Higuchi}
\affiliation{%
Institute for Materials Research, Tohoku University, Miyagi 980-8577, Japan
}%
\author{Naofumi Shimokawa}
\affiliation{%
School of Materials Science, Japan Advanced Institute of Science and Technology, Ishikawa 923-1292, Japan
}%

\date{\today}

\begin{abstract}
Biomembranes, which are mainly composed of neutral and charged lipids, exhibit a large variety of functional structures and dynamics. Here, we report a coarse-grained molecular dynamics (MD) simulation of the phase separation and morphological dynamics in charged lipid bilayer vesicles. The screened long-range electrostatic repulsion among charged head groups delays or inhibits the lateral phase separation in charged vesicles compared with neutral vesicles, suggesting the transition of the phase-separation mechanism from spinodal decomposition to nucleation or homogeneous dispersion. Moreover, the electrostatic repulsion causes morphological changes, such as pore formation, and further transformations into disk, string, and bicelle structures, which are spatiotemporally coupled to the lateral segregation of charged lipids. Based on our coarse-grained MD simulation, we propose a plausible mechanism of pore formation at the molecular level. The pore formation in a charged-lipid-rich domain is initiated by the prior disturbance of the local molecular orientation in the domain.
\end{abstract}

\maketitle

\section{INTRODUCTION}
\label{intro}

Biomembranes, which are formed by self-assembly of various amphiphilic molecules, exhibit a large variety of functional structures and dynamics because these membranes are small, soft, and heterogeneous. There is dynamic compositional heterogeneity in biomembranes, namely, the transient formation of small domains enriched in saturated lipids and cholesterol, known as lipid rafts~\cite{Eggeling,TA}. Raft domains may play important roles in signal transduction and membrane trafficking~\cite{SI,SS,Suzuki}. 
To study the physicochemical mechanism of this lateral membrane heterogeneity, phase separation in multi-component lipid membranes, especially in giant unilamellar vesicles (GUVs) consisting of mixtures of saturated and unsaturated lipids, has attracted a great deal of attention as a model {\it in vitro}. 

Typically, in ternary lipid mixtures composed of saturated lipids, unsaturated lipids, and cholesterol, liquid-ordered ($L_{\mathrm{o}}$) and liquid-disordered ($L_{\mathrm{d}}$) phases coexist~\cite{Keller1,Baumgart,KA}. In particular, the equilibrium conditions for stable phase separation~\cite{Keller2,Feigenson} and the domain-growth dynamics after the sudden transition into coexistence conditions~\cite{Saeki,Yanagisawa2,Keller4} have been extensively investigated, confirming both the equilibrium and dynamic features of phase separation in electrically neutral systems. 
Recently, there has been increasing interest in the effects of long-range electrostatic interactions on the phase separation~\cite{Dimova1,Shimokawa1,Keller3,Dimova2,Himeno1} because plasma membranes~\cite{vanDeenen,Yeung} and some organelles, such as lysosomes~\cite{Dowhan} and mitochondria~\cite{Schlame}, are enriched with charged lipids. In experiments with systems containing charged lipids, charged unsaturated lipids tend to suppress the phase separation compared with neutral lipid systems~\cite{Dimova1,Shimokawa1,Keller3,Dimova2,Himeno1}, whereas charged saturated lipids promote phase separation~\cite{Himeno1}. Several theoretical models~\cite{May,Shimokawa2,Okamoto,Shimokawa3} have also been proposed to explain the electrostatic effects on the lateral phase separation in charged lipid membranes. 

Morphological changes in lipid bilayer membranes are important dynamic phenomena in biomembranes. Examples include the budding in endo- and exocytosis~\cite{MG}, the fission-fusion sequence of vesicular transport~\cite{BBGRB}, and pore formation in autophagy~\cite{JN}. Inspired by the dynamic changes in biomembrane morphology, studies have also investigated the morphological changes in GUVs induced by external stimuli, such as osmotic pressure~\cite{BEH}, addition of surfactant molecules~\cite{Hamada1}, and a synthetic photosensitive amphiphile~\cite{Hamada2}. Some ions or charged biomolecules can also trigger dynamic morphological changes through adsorption onto the lipid monolayers or bilayers via the electrostatic interactions. For example, locally injected HCl drives cristae-like deformation~\cite{Khalifat2008}, BAR (Bin/amphiphysin/Rvs) domain family proteins induce the formation of tubular structures on the membranes~\cite{Saarikangas}, and actin cytoskeletons with myosin motor proteins cause concave deformations of oppositely charged lipid membranes~\cite{Ito2015,Nishigami2016}.

In biomembranes, phase separation and morphological changes in the membrane are closely coupled. Recently, raft domains have been reported to play a critical role in membrane dynamics, such as autophagy and endocytosis, in response to chemical signals~\cite{RS,DCD}.
Using GUVs composed of neutral lipids, coupling between phase separation and morphological dynamics has also been observed. For example, the adhesion between GUVs consisting of inverse cone- and cylinder-shaped lipids~\cite{Sakuma1}, the pore formation in membranes consisting of cone- and cylinder-shaped lipids~\cite{Sakuma2}, and the coupling between the phase separation and deformed membranes by osmotic pressure~\cite{Yanagisawa1,Hamada3} have been reported.
Several studies have focused on the coupling between the phase separation and the morphology in charged lipid membranes. For instance, a positively charged protein, cytochrome c, induces the collapse of domains enriched with cardiolipin (CL$^{(-)}$), which is a negatively charged unsaturated lipid abundant in the inner membrane of mitochondria~\cite{BBGGV}. Even for more physiologically relevant phospholipids with ions, the domains enriched with the unsaturated negatively charged lipid (dioleoylphosphatidylserine; DOPS$^{(-)}$) bud toward the interior of the GUV upon the addition of calcium ions~\cite{Shimokawa1}, and those enriched with the saturated charged lipids (dioleoylphosphatidylglycerol; DOPG$^{(-)}$) undergo pore formation in the membrane by electrostatic repulsion~\cite{Himeno2}.

Although phase separation and the morphologies of charged lipid membranes have been extensively investigated, there are still several questions about the mechanism of charge-specific dynamics. For example, although the domain-coarsening dynamics are frequently examined in neutral membranes~\cite{Saeki,Yanagisawa2,Keller4}, these experiments have not been performed in charged lipid membranes due to the difficulty in preparing experimental systems in which the charge and ion conditions are well-controlled. It is difficult to obtain reliable statistics because the established vesicle-preparation method of electroformation is not applicable to charged systems~\cite{Keller3}.
To reveal the phase-separation dynamics, morphological changes, and the coupling between them in charged lipid membranes, we use a coarse-grained molecular dynamics (MD) simulation~\cite{Markvoort,Zheng} of binary charged and neutral lipid mixtures. To capture such mesoscopic and long-time features of vesicle dynamics, highly coarse-grained model is necessary~\cite{Venturoli,Noguchi}, as we successfully reproduced the experimental morphologies of charged lipid vesicles in our previous work using the present simulation model~\cite{Himeno2}. The coarse-grained MD simulation is suitable for this purpose because we can specify various conditions, such as lipid composition and salt concentration, we can easily investigate the effects of pure electrostatic interactions on a membrane composed of a number of dynamically moving molecules, and we can observe the underlying mechanism at an adequate molecular level.
We investigate the dynamic growth of the charged domain and the coupling between the phase separation and the morphological changes including topological changes. The details of the coarse-grained MD simulation are introduced in Sec.~\ref{methods}. The phase separation and the domain growth are investigated in Secs.~\ref{phase_separation} and ~\ref{coarsening}. The morphological changes induced by electrostatic interactions are presented in Sec.~\ref{morphology}. The implications of our work and comparisons with other studies are discussed in Sec.~\ref{discussion}.

\section{METHODS}
\label{methods}

We extended the coarse-grained model described by Cooke {\it et al.}~\cite{Cooke} to calculate the behavior of charged lipid membranes. This highly coarse-grained model has been used to simulate various mesoscopic behaviors~\cite{Cooke,Guigas}, where the physical properties of the same orders of magnitude with more detailed models have been obtained~\cite{Wang}.
A lipid molecule is represented by one hydrophilic head bead followed by two hydrophobic tail beads. 
The excluded-volume interaction between two beads separated by a distance $r$ is
\begin{equation}
\label{rep}
V_{\mathrm{rep}}(r;b)=\begin{cases}
                  4v\left[ \left( \frac{b}{r} \right)^{12} - \left( \frac{b}{r} \right)^{6} + \frac{1}{4} \right], & r \leq r_{\mathrm{c}}, \\
                  0, & r>r_{\mathrm{c}},
                 \end{cases}
\end{equation}
where $r_{\mathrm{c}}=2^{1/6}b$. $v$ is the unit of energy. 
We chose $b_{\mathrm{head,head}}=b_{\mathrm{head,tail}}=0.95 \sigma$ and $b_{\mathrm{tail,tail}}=\sigma$ to form a stable bilayer, where $\sigma = 7.09$\AA\,is the unit of length corresponding to the cross-sectional diameter of a single lipid molecule.
The potentials for the stretching and bending of bonds between connected beads are expressed as
\begin{equation}
\label{bond}
V_{\mathrm{bond}}(r)=\frac{1}{2}k_{\mathrm{bond}}(r-\sigma)^{2}
\end{equation}
and
\begin{equation}
\label{bend}
V_{\mathrm{bend}}(\theta)=\frac{1}{2}k_{\mathrm{bend}}(1-\cos \theta)^{2},
\end{equation}
where $k_{\mathrm{bond}}=500v/{\sigma^2}$ and $k_{\mathrm{bend}}=60v$ are the bonding strength of connected beads and the bending stiffness of a lipid molecule, respectively.
$\theta$ is the angle between adjacent bond vectors.
The hydrophobic attractive interaction among hydrophobic beads is expressed as
\begin{equation}
\label{attr}
V_{\mathrm{attr}}(r)=\begin{cases}
                -v, & r<r_{\mathrm{c}}, \\
                -v \cos^{2} \left[\frac{\pi(r-r_{\mathrm{c}})}{2 w_{\mathrm{c}}} \right], & r_{\mathrm{c}} \leq r \leq r_{\mathrm{c}}+w_{\mathrm{c}}, \\
                0, & r>r_{\mathrm{c}}+w_{\mathrm{c}},
                 \end{cases}
\end{equation}
where $w_{\mathrm{c}}$ is the cut-off length for the attractive potential. The lipid membrane is in the gel phase when $w_{\mathrm{c}}$ is large, whereas it is in the liquid phase when $w_{\mathrm{c}}$ is small. Such gel and fluid phases observed in the binary lipid bilayer membranes are directly relevant to the liquid-ordered ($L_\mathrm{o}$) and liquid-disordered ($L_\mathrm{d}$) phases, respectively, which are experimentally observed in ternary lipid bilayer membranes composed of unsaturated lipids, saturated lipids, and cholesterols, as both lateral phase separations in the simulation and experiments are recognized by circular domains. Thus, it is possible to determine whether the lipid species is in the gel or liquid phase by controlling $w_{\mathrm{c}}$ in a neutral lipid system~\cite{Cooke}. To represent the charged lipids, we considered the electrostatic interaction among charged head groups in addition to the other interactions. The electrostatic repulsion is described as the Debye-H\"{u}ckel potential
\begin{equation}
\label{elec}
V_{\mathrm{elec}}(r)=v \ell_{\mathrm{B}} z_{1}z_{2}\frac{\exp(-r/\ell_{\mathrm{D}})}{r},
\end{equation}
where $\ell_{\mathrm{B}}=\sigma$ is the Bjerrum length, $z_{1}$ and $z_{2}$ are the valencies of the interacting charged head groups,
and $\ell_{\mathrm{D}}=\sigma \sqrt{\epsilon k_{\mathrm{B}}T/n_{0}e^{2}}$ is the Debye screening length which is related to the bulk salt concentration $n_{0}$ ($\epsilon$, $k_{\mathrm{B}}$, $T$, and $e$ are the dielectric constant of the solution, Boltzmann constant, absolute temperature, and elementary charge, respectively). To capture the behaviors caused by the screened electrostatic interaction qualitatively, $n_{0}$ is set to $1\,\mathrm{M}$, $100\,\mathrm{mM}$, or $10\,\mathrm{mM}$, varying by an order of magnitude around the physiological salt concentration of $100\,\mathrm{mM}$.
Because the typical negatively charged head groups, such as phosphatidylglycerol (PG), phosphatidylserine (PS), phosphatidylinositol (PI), and phosphatidyl acid (PA), have a monovalent charge, we set $z_{1}=z_{2}=-1$. We did not set a cutoff for this screened electrostatic interaction.

Each bead position $\bm{r}_{i}$ obeys the stochastic dynamics described by the Langevin equation
\begin{equation}
\label{langevin}
m\frac{d^{2}\bm{r}_{i}}{dt^{2}}=-\eta\frac{d\bm{r}_{i}}{dt}+\bm{f}^{V}_{i}+\bm{\xi}_{i},
\end{equation}
where $m=1$ and $\eta=1$ are the mass and drag coefficients, respectively. The force $\bm{f}^{V}_{i}$ is calculated from the derivatives of the interaction potentials Eqs.(\ref{rep})--(\ref{elec}). The constant $\tau=\eta\sigma^{2}/v$ is chosen as the unit for the timescale, and the time increment is set at $dt=7.5 \times 10^{-3}\tau$. The Brownian force $\bm{\xi}_{i}$ satisfies the fluctuation-dissipation theorem
\begin{equation}
\label{flu-dis}
\left< \bm{\xi}_{i}(t)\bm{\xi}_{j}(t')\right>=6v\eta\delta_{ij}\delta(t-t').
\end{equation}
In this study, we calculated bilayer vesicles composed of two lipid species to represent the ternary lipid vesicles with gel and liquid domains observed in experiments~\cite{Cooke}. The calculated lipid composition is a binary mixture of A-lipids with or without a monovalent electric charge at the head beads, corresponding to charged or neutral lipids, respectively, and neutral B-lipids. Binary neutral membranes consisting of neutral A- and B-lipids are used as the reference condition for the binary charged membrane consisting of charged A-lipids and neutral B-lipids. The initial state of the calculation is a spherical lipid bilayer consisting of 5000 lipid molecules in total, where A- and B-lipids are mixed homogeneously at a specified ratio, A:B. The total calculation time is set to $t=1.0$--$10.0\times 7500\tau$, during which the phase-separation or morphological dynamics adequately relaxes.

\section{RESULTS}

\subsection{Lateral phase separation}
\label{phase_separation}

	\begin{figure}[ht]
	\includegraphics{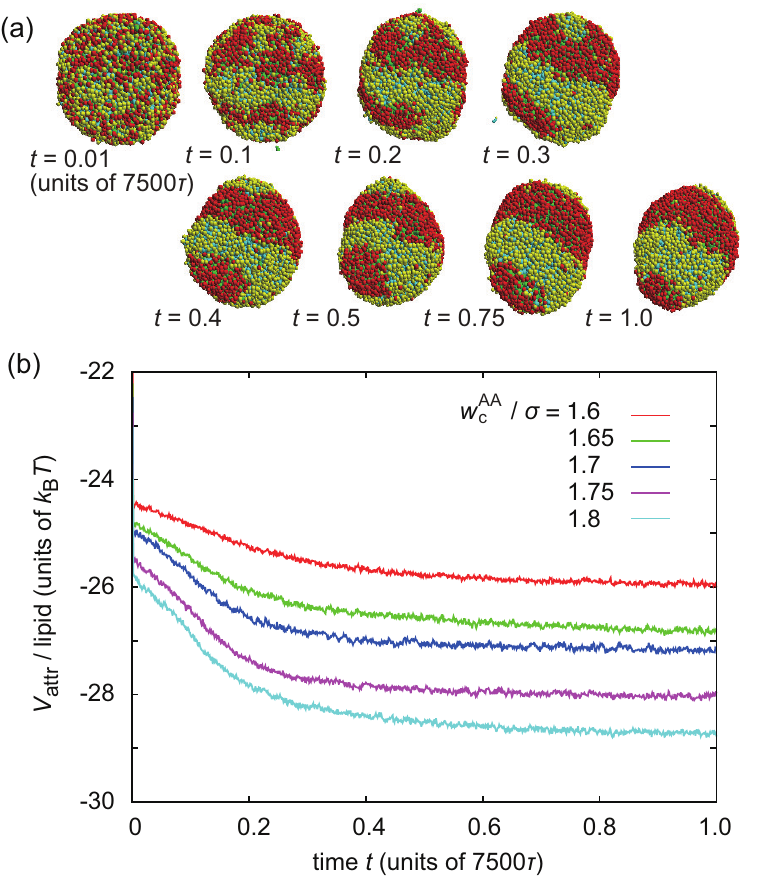}
	\caption{\label{fig:fig1} (a) Typical snapshots of the lateral phase separation of a neutral vesicle consisting of 2500 neutral A-lipids [head, red (dark); tail, green] and 2500 neutral B-lipids [head, yellow (light); tail, cyan]. After rapid relaxation on the order of $t \sim 0.1 \times 7500\tau$, the binary vesicle reaches a quasi-equilibrium state. Attractive interaction parameters are set to $w_\mathrm{c}^\mathrm{AA}/\sigma = 1.7$, $w_\mathrm{c}^\mathrm{BB}/\sigma = 1.7$, and $w_\mathrm{c}^\mathrm{AB}/\sigma = 1.5$. (b) Time evolution of the attractive potential $V_\mathrm{attr}$ per lipid molecule during the lateral phase separation for $t = 1.0 \times 7500\tau$ under various attractive interactions between A-lipids of $w_\mathrm{c}^\mathrm{AA}/\sigma = 1.6$, $1.65$, $1.7$, $1.75$, and $1.8$. Calculations were performed five times for each $w_\mathrm{c}^\mathrm{AA}/\sigma$ value to ensure reproducibility.}
	\end{figure}

	\begin{figure}[ht]
	\includegraphics{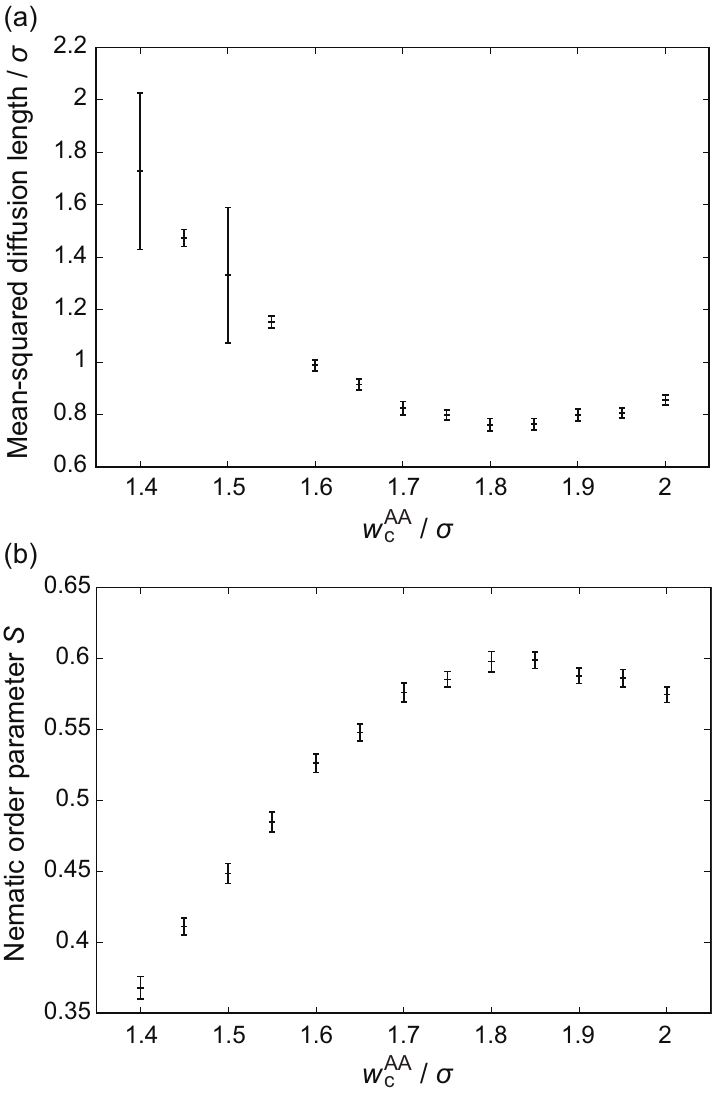}
	\caption{\label{fig:fig2} (a) Mean-squared diffusion length of a lipid in $1000dt$ plotted versus attractive interaction parameter $w_\mathrm{c}^\mathrm{AA}/\sigma$. (b) Nematic order parameter $S$ plotted versus $w_\mathrm{c}^\mathrm{AA}/\sigma$. The values are averaged spatially over a vesicle and temporally over $0.9$--$1.0\times 7500\tau$ for each $w_\mathrm{c}^\mathrm{AA}/\sigma$ in both (a) and (b). Error bars represent the standard deviations. Other interaction parameters are set to $w_\mathrm{c}^\mathrm{BB}/\sigma = 1.7$, and $w_\mathrm{c}^\mathrm{AB}/\sigma = 1.5$.}
	\end{figure}

First, we investigated the dynamic features of lateral phase separation in neutral binary lipid vesicles as the reference behavior for that of charged lipid vesicles. Because both A- and B-lipids are neutral, we set the valencies of their head groups as $z_{\mathrm{A}}=z_{\mathrm{B}}=0$. In other words, force $\bm{f}^{V}$ is obtained from Eqs.(\ref{rep})--(\ref{attr}) without the electrostatic potential in Eq.(\ref{elec}). Figure~\ref{fig:fig1}(a) shows typical snapshots of lateral phase separation in the binary mixture of neutral A-lipids (red or dark) and B-lipids (yellow or light) with A:B = 2500:2500, where the attractive interaction parameters were set to $w_\mathrm{c}^\mathrm{AA}/\sigma = 1.7$, $w_\mathrm{c}^\mathrm{BB}/\sigma = 1.7$, and $w_\mathrm{c}^\mathrm{AB}/\sigma = 1.5$. Immediately after the start of the simulation, the vesicle was in the laterally homogeneous state ($t = 0.01 \times 7500\tau$). The phase separation was gradually induced by the attractive interaction being stronger among the same species than that among different species on a timescale of $t = 0.1 \times 7500\tau$. Figure~\ref{fig:fig1}(b) shows the attractive potential, $V_\mathrm{attr}$, per lipid molecule during the phase separation for attraction parameters $w_\mathrm{c}^\mathrm{AA}/\sigma = 1.6$, $1.65$, $1.7$, $1.75$, and $1.8$, where the bilayer membrane phase ranged from fluid to gel~\cite{Cooke}. We quantitatively confirmed the transition between liquid and gel phases induced by the attractive interaction parameter $w_\mathrm{c}^\mathrm{AA}$ at $w_\mathrm{c}^\mathrm{AA}/\sigma = 1.7$, based on the calculations of the lipid diffusivity and nematic order parameter in the vesicle. Figure~\ref{fig:fig2}(a) shows the mean-squared diffusion length of a lipid in $1000dt$ plotted versus $w_\mathrm{c}^\mathrm{AA}/\sigma$. Although a smaller $w_\mathrm{c}^\mathrm{AA}$ is accompanied by a larger error resulting from spike noises caused by frequent dropouts of lipids, the diffusion length decreased according to the increase in $w_\mathrm{c}^\mathrm{AA}$, and reached a plateau for $w_\mathrm{c}^\mathrm{AA}/\sigma \geq 1.7$. We also calculated the nematic order parameter by $S=\frac{1}{2}\left<3\bm{v}_{\mathrm{b},i}\cdot\bm{v}_{\mathrm{b},j}-1\right>_{i,j}$, where $\bm{v}_{\mathrm{b},i}$ is the unit bond vector of tail beads of the $i$th lipid and $\left<\,\right>_{i,j}$ represents the average over all the combinations of lipid pairs [Fig.~\ref{fig:fig2}(b)]. The nematic order parameter also shows a clear plateau at $w_\mathrm{c}^\mathrm{AA}/\sigma \geq 1.7$, indicating the most ordered state with the lowest diffusivity in this parameter range. These two features in Figs. 2(a) and 2(b) characterize the transitional behavior of lipid bilayer membranes between gel and liquid phases in our simulation. Despite the phase transition around $w_\mathrm{c}^\mathrm{AA}/\sigma = 1.7$, even though the values of $V_\mathrm{attr}$ were slightly different according to $w_\mathrm{c}^\mathrm{AA}$ in Fig.~\ref{fig:fig1}(b), the vesicles exhibited the same qualitative feature in their dynamics, namely, a decrease in potential on a timescale of $t \sim 0.1 \times 7500\tau$. After this rapid relaxation into phase-separated states, the vesicles reached quasi-equilibrium states, identified by the flattening of the temporal changes in $V_\mathrm{attr}$. Complete phase separation with two macro domains is reached stochastically on much longer timescales than the practical calculation time in this study because of the much slower diffusion of large domains compared with single molecule. Because the other kinds of energies $V_\mathrm{rep}$, $V_\mathrm{bond}$, and $V_\mathrm{bend}$ did not exhibit any clear changes during the phase separation, the change in attractive interaction dominated the lateral phase separation in neutral lipid vesicles in both the gel and liquid states. 

	\begin{figure*}[!t]
	\includegraphics{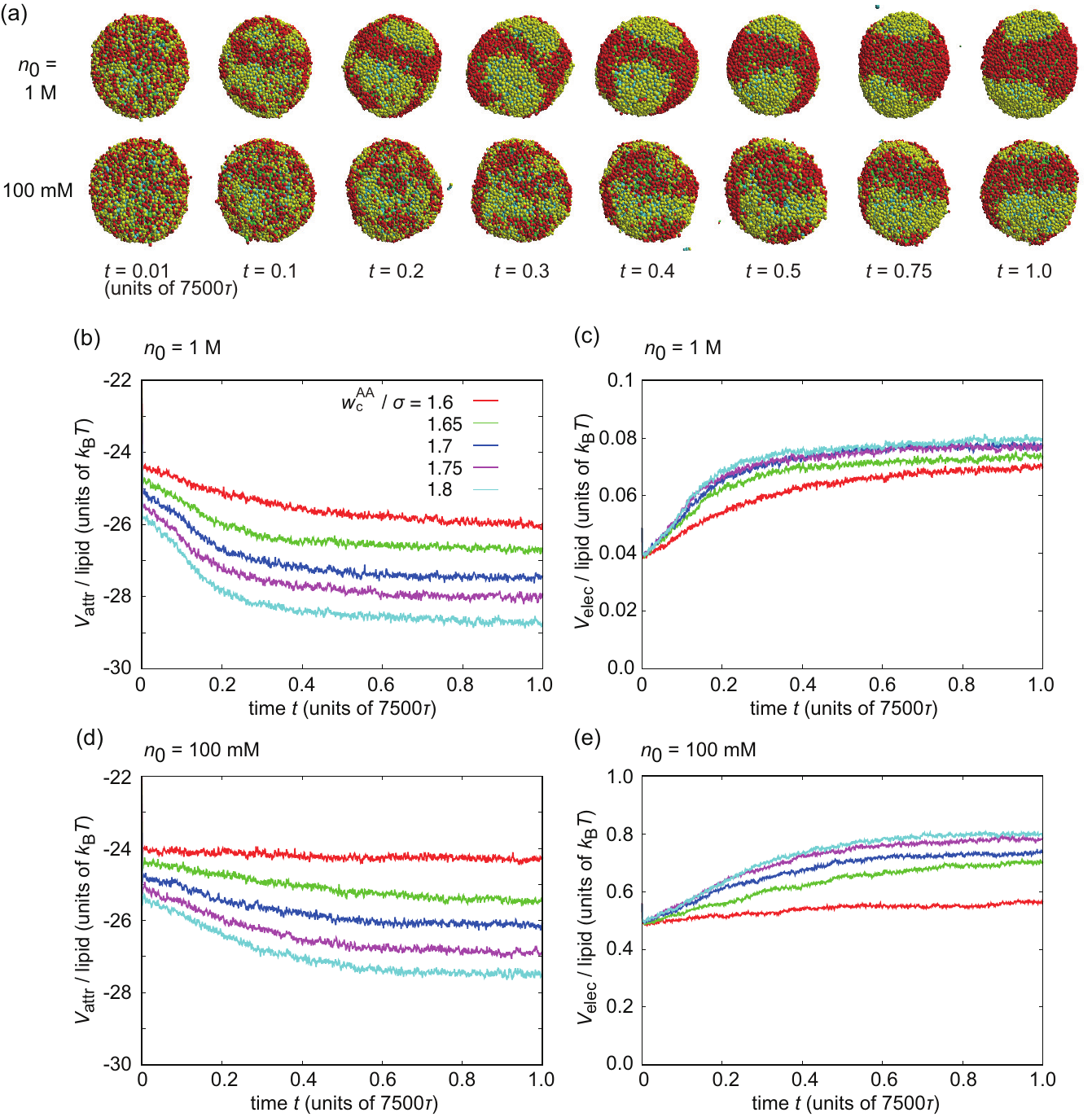}
	\caption{\label{fig:fig3} (a) Typical snapshots of the lateral phase separation of a charged vesicle consisting of 2500 charged A-lipids [head, red (dark); tail, green] and 2500 neutral B-lipids [head, yellow (light); tail, cyan] at salt concentrations $n_{0}=1\,\mathrm{M}$ and $100\,\mathrm{mM}$. Attractive interaction parameters are set to $w_\mathrm{c}^\mathrm{AA}/\sigma = 1.7$, $w_\mathrm{c}^\mathrm{BB}/\sigma = 1.7$, and $w_\mathrm{c}^\mathrm{AB}/\sigma = 1.5$ for both salt conditions. (b) Attractive and (c) electrostatic repulsive potentials $V_\mathrm{attr}$ and $V_\mathrm{elec}$, respectively, for $t = 1.0 \times 7500\tau$ under various attractive interactions between A-lipids of $w_\mathrm{c}^\mathrm{AA}/\sigma = 1.6$, $1.65$, $1.7$, $1.75$, and $1.8$ at $n_{0}=1\,\mathrm{M}$. (d) Attractive and (e) electrostatic repulsive potentials at $n_{0}=100\,\mathrm{mM}$. Calculations were performed three times for each $n_0$ and $w_\mathrm{c}^\mathrm{AA}/\sigma$ value to ensure reproducibility.}
	\end{figure*}

Next, we considered a negatively charged vesicle consisting of negatively charged A-lipids and neutral B-lipids to examine the effect of long-range electrostatic repulsion on the lateral phase separation. The valencies of the head groups of A- and B-lipids were set as $z_{\mathrm{A}}=-1$ and $z_{\mathrm{B}}=0$, respectively. 
Figure~\ref{fig:fig3}(a) shows typical snapshots of the lateral phase separation of vesicles composed of 2500 charged A-lipids and 2500 neutral B-lipids under different salt concentrations. All the parameters except for the valency of the A-lipids were  the same as in the calculation for the neutral vesicles (Fig.~\ref{fig:fig1}). 
	\begin{figure}[!t]
	\includegraphics{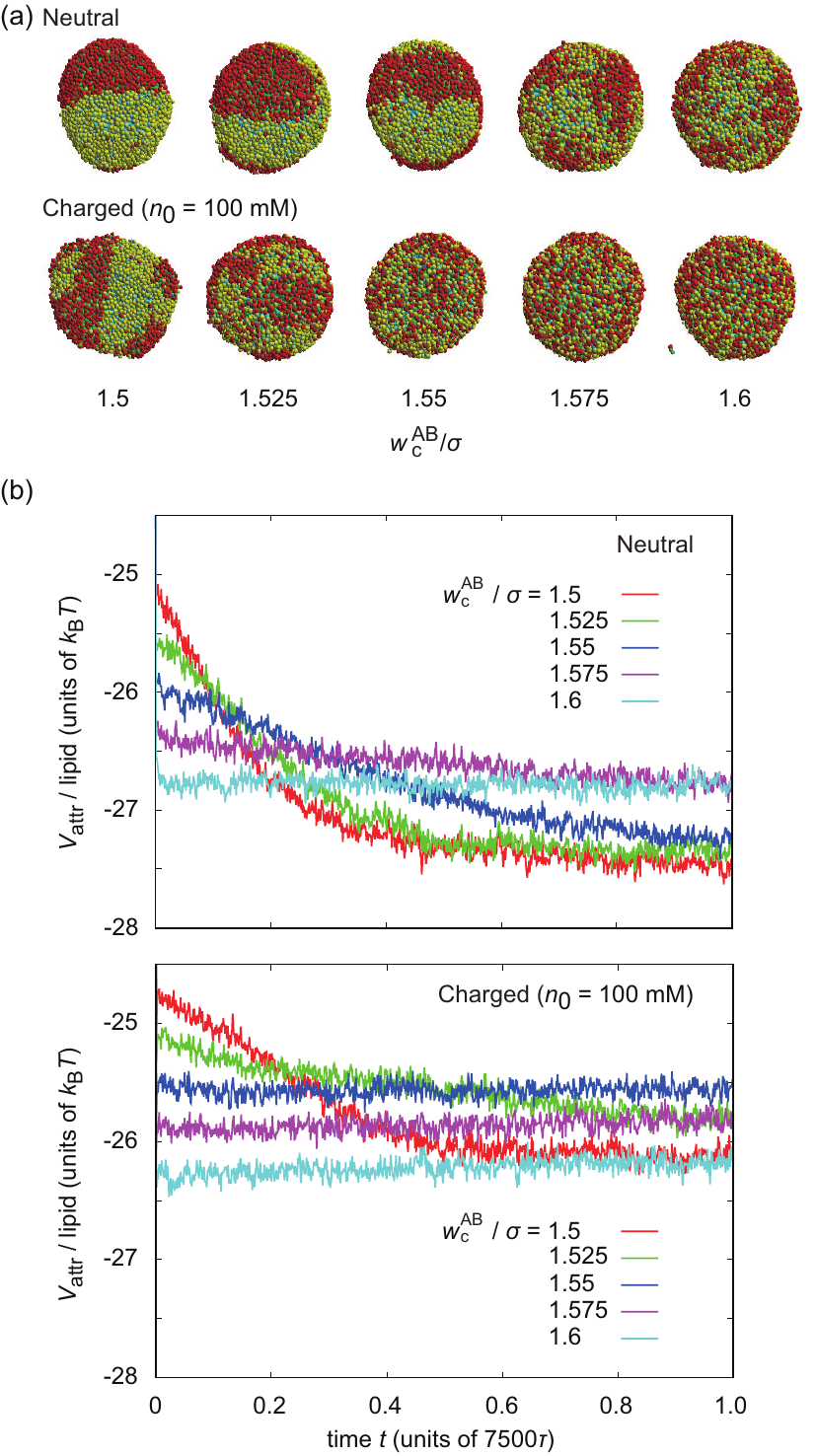}
	\caption{\label{fig:critical} (a) Typical snapshots of the lateral phase separation of neutral (upper) and charged (at $n_0=100\,\mathrm{mM}$, lower) vesicles for $w_\mathrm{c}^\mathrm{AB}/\sigma=1.5$, $1.525$, $1.55$, $1.575$, and $1.6$. Other conditions are set to $w_\mathrm{c}^\mathrm{AA}/\sigma=w_\mathrm{c}^\mathrm{BB}/\sigma=1.7$, A:B = 2500:2500. (b) Corresponding attractive potentials $V_\mathrm{attr}$. Calculations were performed three times for each charge condition and $w_\mathrm{c}^\mathrm{AB}/\sigma$ value to ensure reproducibility.} 
	\end{figure}
For charged lipids at a higher salt concentration of $n_{0}=1\,\mathrm{M}$ the vesicles exhibited dynamics similar to those of the neutral vesicles, whereas at a lower salt concentration of $n_{0}=100\,\mathrm{mM}$ they exhibited clearly slower phase-separation dynamics. The attractive potential $V_\mathrm{attr}$ of the charged lipid vesicles at $n_{0}=1\,\mathrm{M}$ (Fig.~\ref{fig:fig3}(b)) was also quantitatively the same as that of the neutral lipids (Fig.~\ref{fig:fig1}(b)), although the electrostatic repulsive potential, $V_\mathrm{elec}$, increased with the phase-separation process (Fig.~\ref{fig:fig3}(c)). Thus, the electrostatic interaction among charged lipids with $z_{\mathrm{A}}=-1$ was negligible at $n_{0}=1\,\mathrm{M}$, and the increase in $V_\mathrm{elec}$ was two orders of magnitude lower than the decrease in $V_\mathrm{attr}$. However, at $n_{0}=100\,\mathrm{mM}$, the phase-separation dynamics were affected by the less-screened electrostatic repulsion as the decrease in $V_\mathrm{attr}$ became slower and $V_\mathrm{elec}$ increased by an order of magnitude compared with that at $n_{0}=1\,\mathrm{M}$ (Fig.~\ref{fig:fig3}(d) and ~\ref{fig:fig3}(e)).

We compared the static phase-separated states of neutral and charged vesicles consisting of 2500 neutral or charged A-lipids and 2500 neutral B-lipids. We focused on the $w_\mathrm{c}^\mathrm{AB}$ dependence of the static states because the attractive parameter $w_\mathrm{c}^\mathrm{AB}$ governs the degree of mixing, which provides quantitative information about susceptibility of the phase separation in neutral and charged vesicles. As shown in Fig.~\ref{fig:critical}(a), there was a large difference in the the critical attractive parameter, $w_\mathrm{c}^\mathrm{AB}$, for the lateral phase separation. The corresponding changes in the attractive interaction potential, $V_\mathrm{attr}$, shown in Fig.~\ref{fig:critical}(b) clearly indicate the transition point of $w_\mathrm{c}^\mathrm{AB}$, which coincides with the constant interaction potential over $t=1.0\times7500\tau$. The plateau in $V_\mathrm{attr}$ occurred at $w_\mathrm{c}^\mathrm{AB}/\sigma=1.6$ for neutral vesicles, but at $w_\mathrm{c}^\mathrm{AB}/\sigma=1.55$ for charged vesicles at $n_0=100\,\mathrm{mM}$. This inhibition of the lateral phase separation by the electrostatic repulsion among charged lipids qualitatively agrees with previous experiments~\cite{Dimova1,Shimokawa1,Keller3,Dimova2,Himeno1} and theoretical calculations based on the free energy of charged membranes~\cite{Shimokawa1,May}.

\subsection{Coarsening dynamics}
\label{coarsening}
	
	\begin{figure}[ht]
	\includegraphics{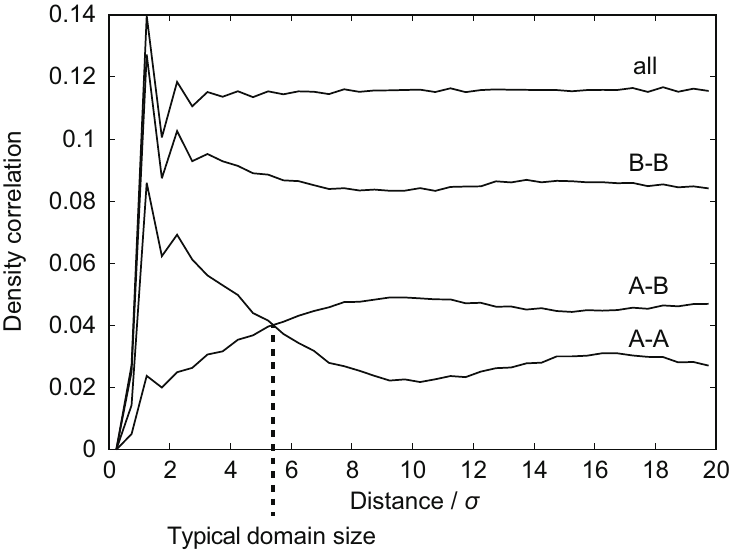}
	\caption{\label{fig:pair} Density correlation functions between lipid species. The intersection between the (A-A) correlation curve and (A-B) correlation curve is the typical domain size of A-lipid-rich domains. Typical conditions, where $w_\mathrm{c}^\mathrm{AA}/\sigma=w_\mathrm{c}^\mathrm{BB}/\sigma=1.7$, $w_\mathrm{c}^\mathrm{AB}/\sigma=1.5$, A:B=1500:3500, and $t=1.0\times 7500\tau$, are used as an example.}
	\end{figure}

	\begin{figure}[ht]
	\includegraphics{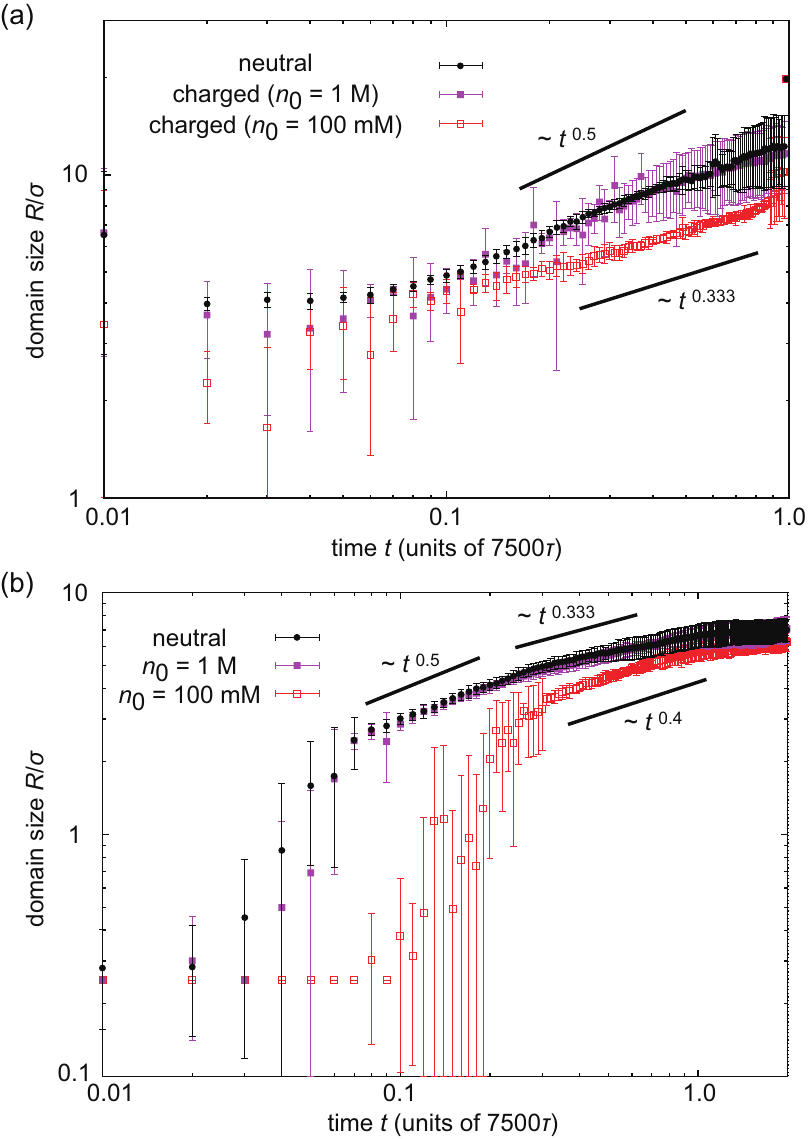}
	\caption{\label{fig:coarsening} (a) Time evolution of domain size for neutral vesicles consisting of neutral A- and B-lipids (black filled circles) and charged vesicles consisting of charged A-lipids and neutral B-lipids at $n_{0}=1\,\mathrm{M}$ (purple filled squares) and $n_{0}=100,\mathrm{mM}$ (red open squares). Lipid composition of A:B = 2500:2500. Error bars represent the standard deviation at each time. Calculations were performed five times for neutral and charged lipids at $n_{0}=1\,\mathrm{M}$, and three times for charged lipids at $n_{0}=100\,\mathrm{mM}$. Slopes for $t^{0.5}$ and $t^{0.333}$ in the double logarithmic axis are shown as visual guides. (b) Lipid composition of A:B = 1500:3500 in the same manner in (a). Calculations were performed twenty times for neutral, and ten times for charged lipids at $n_{0}=1\,\mathrm{M}$ and $100\,\mathrm{mM}$. Slopes for $t^{0.5}$, $t^{0.4}$, and $t^{0.333}$ are shown as visual guides.}
	\end{figure}

The dynamics of phase separation in lipid vesicles has been extensively studied experimentally and theoretically in terms of a unified view of scaling behaviors of the domain coarsening, where the domain size $R(t)$ is proportional to $t^\alpha$~\cite{Keller4,Laradji} and $\alpha$ is the domain-growth exponent. Thus, we analyzed $R(t)$ for neutral and charged lipid vesicles. To define the typical domain size, the density correlation function averaged over the vesicle was calculated. Figure~\ref{fig:pair} shows the spatial distance at which the density autocorrelation function of A-lipids (A-A) intersects with the density cross-correlation function between A- and B-lipids (A-B); this distance is defined here as the typical size of A-lipid-rich domains. Figure~\ref{fig:coarsening} shows the different coarsening dynamics in the neutral lipid vesicles, charged lipid vesicles for $n_{0}=1\,\mathrm{M}$, and charged lipid vesicles for $n_{0}=100\,\mathrm{mM}$ at the lipid compositions of A:B = 2500:2500 (Fig.~\ref{fig:coarsening}(a)) and 1500:3500 (Fig.~\ref{fig:coarsening}(b)). At A:B = 2500:2500, the domain-growth exponents of neutral A-lipid-rich domains and charged A-lipid-rich domains at $n_{0}=1\,\mathrm{M}$ were both $\alpha \sim 0.5$ (Fig.~\ref{fig:coarsening}(a)), which is generally seen in the coarsening process caused by domain-area deformation driven by line tension~\cite{Saeki,Yanagisawa2,Keller4}. Under these conditions, the observed phase separation occurred by spinodal decomposition with percolated domain. This bicontinuous structure with a clear domain edge was deformed over time by the line tension between the A- and B-lipid-rich phases, which originated from the different attractive interactions between the same and different species. In contrast, the coarsening of the charged A-lipid-rich domains at $n_{0}=100\,\mathrm{mM}$ showed slower dynamics and $\alpha \sim 1/3$, which is generally seen during evaporation condensation (Ostwald ripening), where the growth exponent of $1/3$ is referred to as the Lifshitz--Slyozov--Wagner model~\cite{Lifshitz,Wagner}, or during diffusion-limited collision and coalescence of small liquid domains~\cite{Laradji}. Figure~\ref{fig:fig3}(a) at $n_{0}=100\,\mathrm{mM}$ exemplifies the different slow coarsening dynamics in the binary vesicles in the presence of relatively strong long-range repulsion. The domain edge was blurred in this case, and a number of small domains randomly came together rather than a single percolated domain undergoing tension-driven deformation .

In addition, we examined the lipid composition of A:B = 1500:3500 (Fig.~\ref{fig:coarsening}(b)), for which the area fraction of charged A-lipids was almost $0.3$, because the coarsening exponent $\alpha$ depends on the area fraction of the phase-separated domains~\cite{Keller4}. The coarsening dynamics of the neutral lipid vesicles and charged lipid vesicles at $n_{0}=1\,\mathrm{M}$ were again similar. After the initial domain formation with a domain size of more than 2 or 3 lipids, the domain-growth exponent was $\alpha \sim 0.5$ as in the case of A:B = 2500:2500; however, $\alpha$ decreased to $\sim 0.3$ or slightly less, probably due to the small number of A-lipids and A-lipid-rich domains. The small number of lipids formed small circular domains rather than a percolated domain, resulting in the collision and coalescence of the liquid domains with an exponent of $1/3$. The coarsening dynamics of charged lipid vesicles at $n_{0}=100\,\mathrm{mM}$ were also an order of magnitude slower than those under neutral or well-screened charged conditions. Interestingly, the vesicles remained for a time during the initial homogeneous state and then suddenly started coarsening, with an ill-defined exponent of $\alpha \sim 0.4$ or less.

\subsection{Morphological changes}
\label{morphology}

	\begin{figure}[!ht]
	\includegraphics{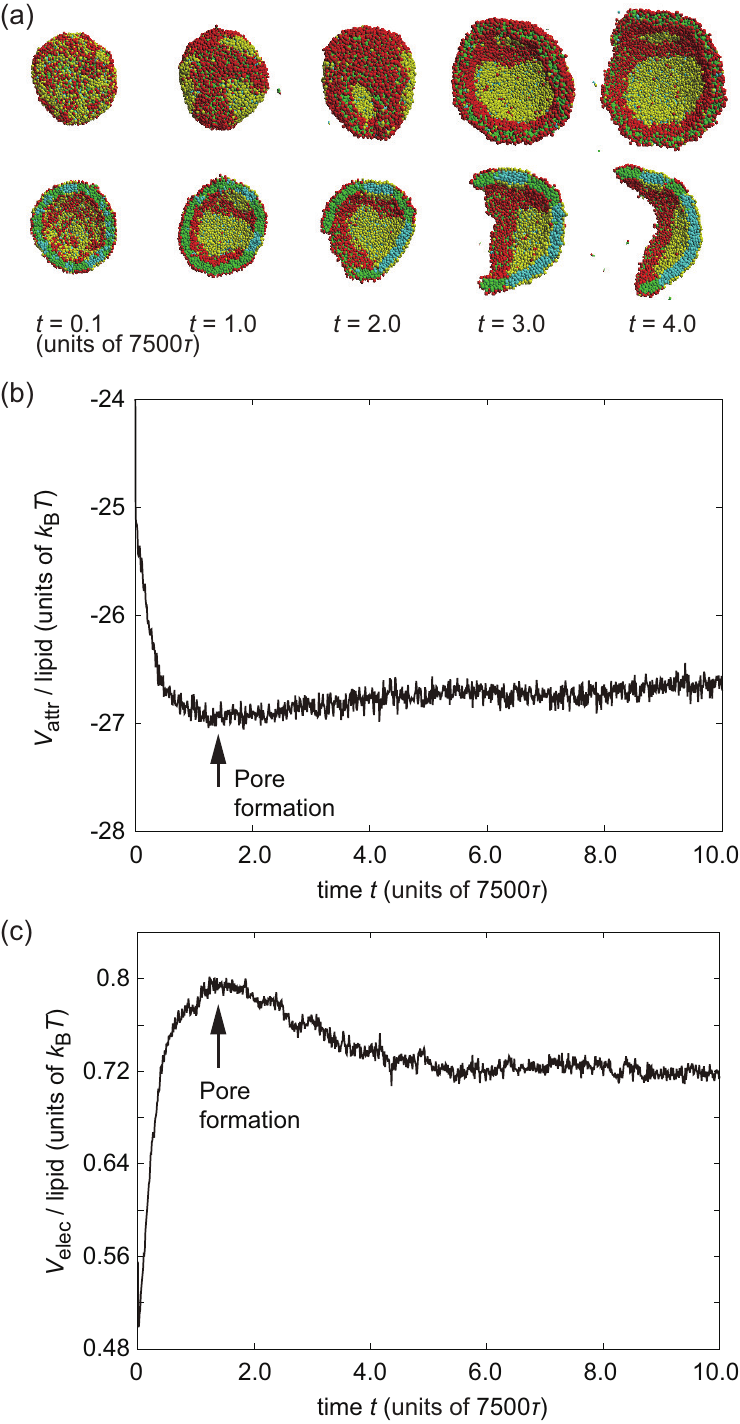}
	\caption{\label{fig:pore_enr} (a) Typical snapshots of the morphological changes of a binary charged vesicle composed of 2500 charged A-lipids (head, red (dark); tail, green) and 2500 neutral B-lipids (head, yellow (light); tail, cyan) at $n_{0} = 100\,\mathrm{mM}$. $w_\mathrm{c}^\mathrm{AA}/\sigma = 1.75$, $w_\mathrm{c}^\mathrm{BB}/\sigma = 1.7$, and $w_\mathrm{c}^\mathrm{AB}/\sigma = 1.5$. The upper and lower image series respectively show surface and cross-sectional images of the same vesicle. Typically, the morphological changes occur an order of magnitude slower ($t \sim 1.0 \times 7500 \tau$) than the lateral phase separation ($t \sim 0.1 \times 7500 \tau$). Time evolution of the (b) attractive and (c) electrostatic repulsive potentials $V_\mathrm{attr}$ and $V_\mathrm{elec}$ per lipid molecule, respectively. The calculation was performed once and showed plausible results (see Fig.~\ref{fig:morph_diagdam}).}
	\end{figure}

In this section, we turn our focus to the morphological dynamics of the charged lipid vesicles. In the previous section, we calculated the phase-separation dynamics on a timescale of $t=1.0 \times 7500 \tau$. However, when we continued the calculation until $t=10.0 \times 7500 \tau$, we observed the morphological changes in the charged lipid vesicles depending on the salt concentration $n_{0}$. In contrast, we did not observe morphological changes within $t=10.0 \times 7500 \tau$ for neutral lipid vesicles (data not shown). Figure~\ref{fig:pore_enr}(a) shows typical snapshots of the morphological changes, including the topological change from spherical to disk-shaped, which were observed during the longer time calculation at $n_{0}=100\,\mathrm{mM}$. The lipid composition was set to A:B = 2500:2500, and the interaction parameters were set to $w_\mathrm{c}^\mathrm{AA}/\sigma = 1.75$, $w_\mathrm{c}^\mathrm{BB}/\sigma = 1.7$, and $w_\mathrm{c}^\mathrm{AB}/\sigma = 1.5$. After the relaxation of the lateral phase separation for $t = 1.0 \times 7500\tau$ (second image in Fig.~\ref{fig:pore_enr}(a)), a pore was formed at a charged-lipid-rich domain (red or dark) in the membrane at around $t = 1.3 \times 7500\tau$. The pore diameter increased with time and finally the vesicle transformed into a disk shape. The charged lipids were segregated at the edge of the disk, whereas the neutral lipids formed a planar bilayer membrane surrounded by the charged lipids (last image in Fig.~\ref{fig:pore_enr}(a)). During the pore formation, the almost relaxed attractive potential began to increase (Fig.~\ref{fig:pore_enr}(b)), whereas the increased electrostatic repulsive potential started to decrease (Fig.~\ref{fig:pore_enr}(c)). After the onset of pore formation at $t \sim 1.3 \times 7500\tau$, these potentials changed as the vesicle morphology relaxed to the flat disk shape in $t \sim 7.0 \times 7500\tau$. Although the potentials slightly changed even after $t \sim 7.0 \times 7500\tau$, due to the gradual dropouts of lipid molecules from the structure, we confirmed that the disk shape was kept up to $t = 20.0 \times 7500\tau$. We thus regarded the morphologies at $t = 10.0 \times 7500\tau$ as the static states of the pre-assembled vesicles. Note that such states observed in experiments have been treated as practically static because of the much slower timescale of complete ``melting'' of the structure.

	\begin{figure}[!hb]
	\includegraphics{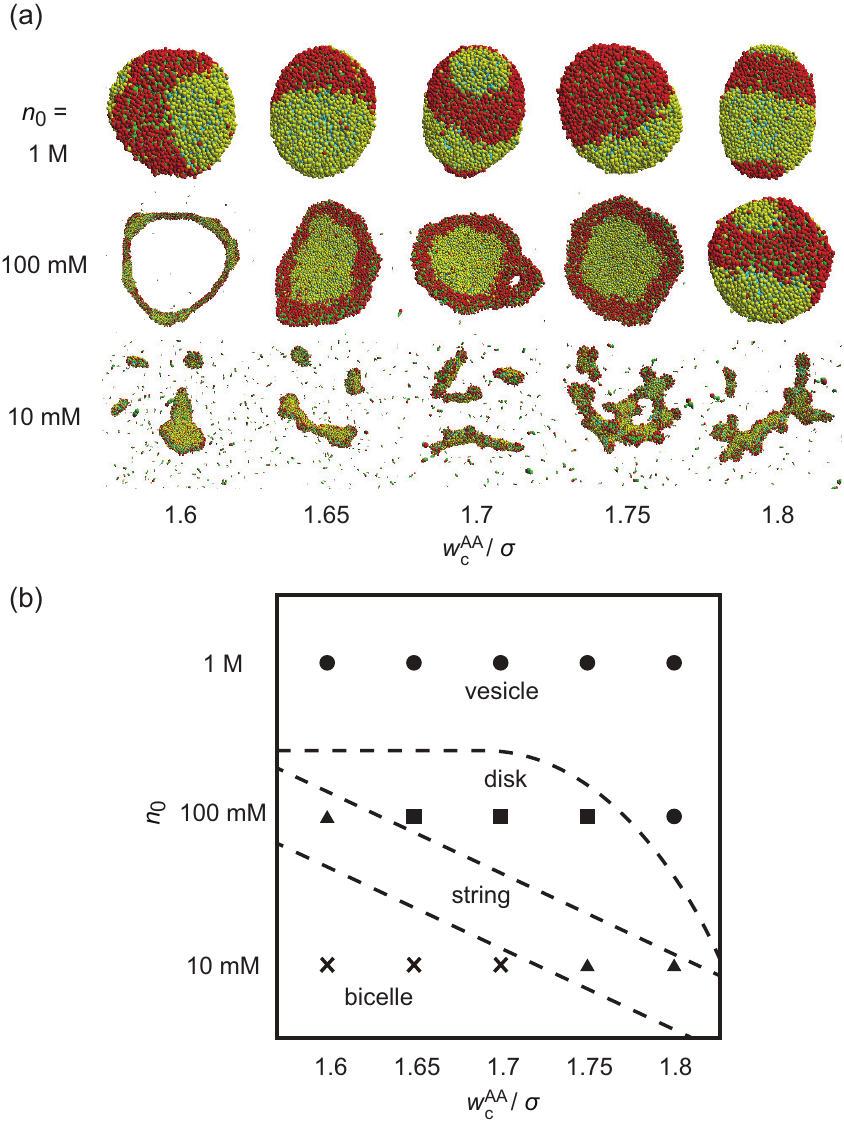}
	\caption{\label{fig:morph_diagdam} (a) Typical morphologies for various salt conditions and attractive interactions between A-lipids. $n_{0}$ is from $10\,\mathrm{mM}$ to $1\,\mathrm{M}$ and $w_\mathrm{c}^\mathrm{AA}/\sigma$ is from $1.6$ to $1.8$. $w_\mathrm{c}^\mathrm{BB}/\sigma$ and $w_\mathrm{c}^\mathrm{AB}/\sigma$ were fixed as $1.7$ and $1.5$, respectively. (b) Phase diagram of the charged vesicle morphologies. We defined four types of morphologies; spherical vesicle (circle), disk (square), string (triangle), and bicelle (cross). The dashed lines are visual guides indicating the phase boundaries.}
	\end{figure}

Finally, we determined the phase diagram of the vesicle morphology for various salt concentrations and attractive interaction parameters between A-lipids (Fig.~\ref{fig:morph_diagdam}). The salt concentration, $n_{0}$, was varied from $1\,\mathrm{M}$ to $10\,\mathrm{mM}$, ranging from almost neutral conditions with a corresponding Debye screening length of $0.43\sigma$ ($3.05\,$\AA) to highly repulsive conditions with $4.3\sigma$ ($30.5\,$\AA). The attraction parameter for A-lipids, $w_\mathrm{c}^\mathrm{AA}/\sigma$, was varied from $1.6$ (liquid state) to $1.8$ (gel state). In this wide parameter region, Fig.~\ref{fig:morph_diagdam}(a) shows the typical morphologies of charged lipid vesicles coupled to the lateral phase separation observed at $t=10.0\times7500\tau$, when the vesicles reach static states as shown in Figs.~\ref{fig:pore_enr}(b) and ~\ref{fig:pore_enr}(c). At $n_{0}=1\,\mathrm{M}$, the lipid vesicles remained spherical and macro-phase separation was observed for all the values of $w_\mathrm{c}^\mathrm{AA}$. At $n_{0}=100\,\mathrm{mM}$, the vesicles exhibited various morphologies depending on $w_\mathrm{c}^\mathrm{AA}$. For $w_\mathrm{c}^\mathrm{AA}/\sigma = 1.6$, the initially spherical vesicles became string-shaped. For $w_\mathrm{c}^\mathrm{AA}/\sigma = 1.65$, $1.7$, and $1.75$, the vesicles became disk-shaped, and for $w_\mathrm{c}^\mathrm{AA}/\sigma = 1.8$, they remained spherical. At $n_{0}=10\,\mathrm{mM}$, where the electrostatic repulsion was substantially stronger than the attractive interaction among lipid molecules, the self-assembled structures were no longer maintained. From $w_\mathrm{c}^\mathrm{AA}/\sigma = 1.6$ to $1.7$ or $1.75$, the vesicles broke into small aggregations, called bicelles, whereas at $w_\mathrm{c}^\mathrm{AA}/\sigma = 1.8$ the vesicles adopted a spiky string morphology. When the transitions to disk, string, or bicelle occurred, the charged lipids assembled at the edge of these static structures. Figure~\ref{fig:morph_diagdam}(b) shows a phase diagram summarizing these static shapes.

\section{DISCUSSION}
\label{discussion}

In this study, we investigated the effects of screened but still long-range electrostatic repulsion on the dynamic behaviors of lipid bilayer vesicles, especially the lateral-phase-separation dynamics and morphological changes. 
After a temperature quench below the miscibility transition temperature, the phase-separation dynamics of a homogeneous neutral lipid membrane are nucleation growth from a metastable state or spinodal decomposition from an unstable state. During the nucleation process, there is an effective energy barrier against the localization of a lipid species, owing to factors such as the mixing entropy of lipids and the line tension of domains. In contrast, during spinodal decomposition, a binary lipid mixture immediately segregates into two coexisting phases. The occurrence of these two instability processes generally depends on the level of the parameter quench at the onset of the phase separation and the lipid composition of the vesicle, namely, whether the system state is located below the spinodal line or between the binodal and spinodal lines in the two-parameter field~\cite{Keller4,Lipowsky}. For example, Veatch and Keller reported the instability of stripe-shaped domains with only neutral lipids, suggesting spinodal decomposition in a vesicle of (dioleoylphosphatidylcholine; DOPC):(dipalmitoylphosphatidylcholine; DPPC) (1:1) and 35\% cholesterol caused by a temperature quench near the critical miscibility temperature~\cite{Keller2}. After the onset of the phase separation by nucleation or spinodal decomposition, the emergent domains grow further by evaporation condensation (Ostwalt ripening), collision and coalescence by diffusion, and domain deformation by line tension, depending on the area fraction of the domain-forming lipids~\cite{Keller4}. These domain growth processes at this stage are called ``coarsening''.

	\begin{figure*}[!ht]
	\includegraphics{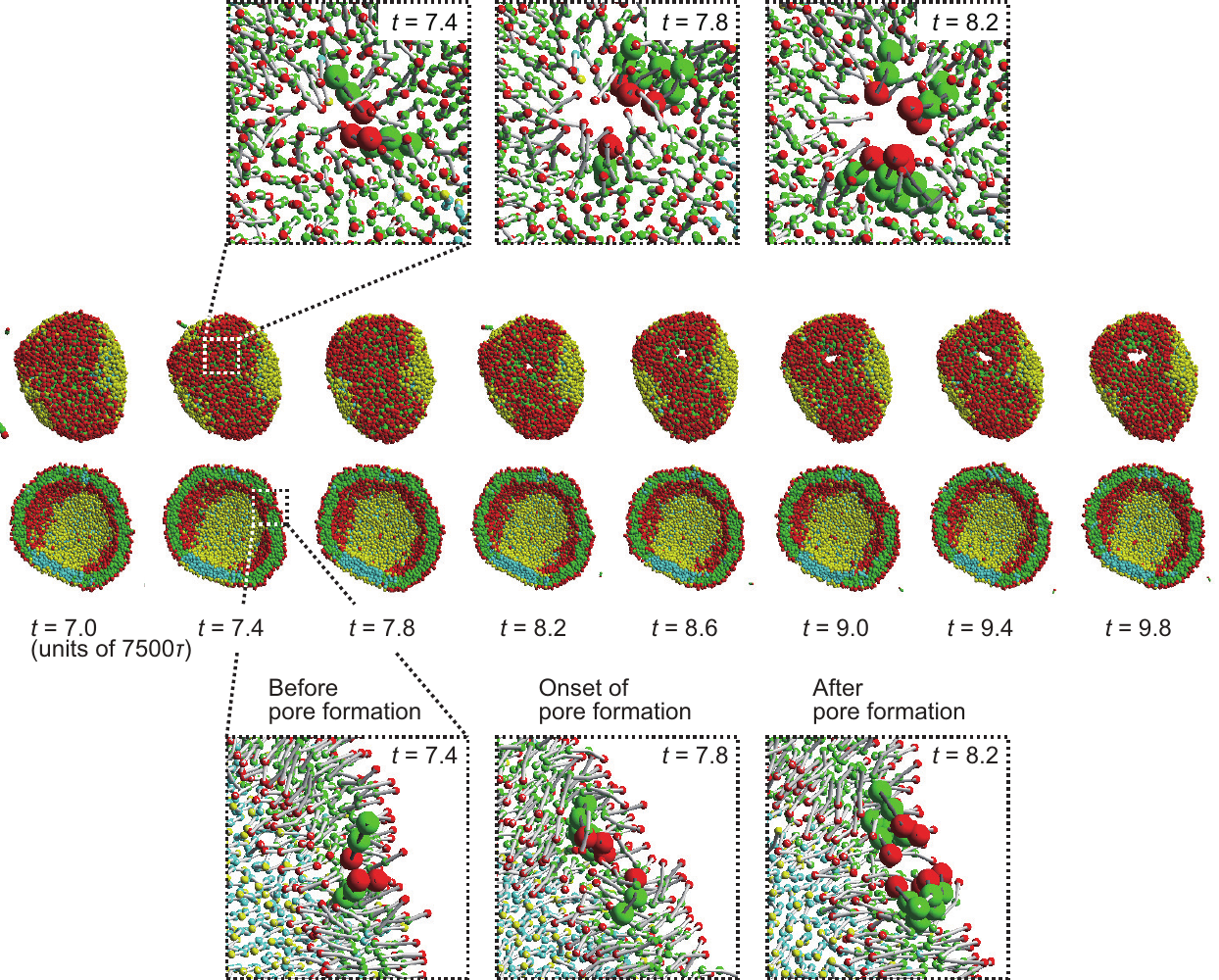}
	\caption{\label{fig:detail} Example of the molecular mechanism of pore formation. The vesicle is composed of 2500 charged A-lipids (head, red (dark); tail, green) and 2500 neutral B-lipids (head, yellow (light); tail, cyan) at $n_{0} = 100\,\mathrm{mM}$. $w_\mathrm{c}^\mathrm{AA}/\sigma = w_\mathrm{c}^\mathrm{BB}/\sigma = 1.7$ and $w_\mathrm{c}^\mathrm{AB}/\sigma = 1.5$. Sequential snapshots in the middle show the surface and cross-sectional images taken at $t=7.0\times7500\tau$ to $9.8\times7500\tau$. Dotted squares contain magnified images of the pore-forming site at $t=7.4\times7500\tau$ (before), $7.8\times7500\tau$ (onset), and $8.2\times7500\tau$ (after), for surface (upper) and cross-sectional (lower) views of the bilayer membrane. Embedded lipids and the surrounding lipids are displayed with spheres of the original size and small spheres, respectively.}
	\end{figure*}

The present numerical calculations revealed that the attraction parameters, the lipid composition, and the electrostatic long-range repulsion play important roles in the coarsening process and the membrane instability. Although it is difficult to observe the exact onset of the phase separation owing to the short length and fast timescale, our MD simulation can shed light on this phenomenon. In the coarsening dynamics of charged lipid membranes at $n_{0}=100\,\mathrm{mM}$ for the lipid composition A:B = 2500:2500, the lateral structure of the membrane remained homogeneous for a certain duration, whereas the neutral lipid membranes or charged lipid membranes at $n_{0}=1\,\mathrm{M}$ immediately began the phase separation (Fig.~\ref{fig:coarsening}(b)), even though the parameter conditions were the same. This result strongly indicates that the dynamics of the charged lipid membranes at the onset of the phase separation are not spinodal decomposition, but instead nucleation growth starting from a metastable state with an energy barrier, probably originating from the electrostatic repulsive force among charged lipids, that needs an overcoming time $t \sim 0.1 \times 7500\tau$. 
During late phase separation, the domain growth in charged membranes was much smaller than that in neutral membranes (Fig.~\ref{fig:coarsening}), and growth exponent $\alpha$ values were also different from those in neutral membranes. Pataraia {\it et al.} previously observed transient small circular domains or bicontinuous domains even after 20 min from vesicle preparation of ternary charged vesicles composed of negatively charged dioleoylphosphatidylglycerol (DOPG$^{(-)}$), egg sphingomyelin (eSM), and cholesterol, indicating the slower coarsening dynamics. This finding agrees well with our numerical calculations and implies the underlying general physicochemical effects on the phase-separation process of electrostatic interaction among charged lipids. However, the details of the coarsening dynamics, such as growth exponent $\alpha$ of charged lipid membranes, remains less well understood. In this study, we observed an ill-defined exponent $\alpha\sim0.4$ for A:B = 1500:3500 [Fig.~\ref{fig:coarsening}(b)]. Two-dimensional charged membranes can be exempted from electroneutrality because their counterions can escape into the three-dimensional bulk, and so the coarsening dynamics in charged membranes might be a unique feature of a two-dimensional electrolyte solution. Therefore, examining $\alpha$ for charged lipid membranes in other experimental or theoretical works would be an interesting step toward understanding the dynamic heterogeneity in charged membrane systems.

Li {\it et al.} recently reported the first systematic theoretical study of the equilibrium conformational changes caused by the electrostatic interactions among surface lipids by calculating the minimum bending and Coulombic energies~\cite{Li}. Although they assumed a fixed uniform surface charge distribution, our model can treat a more plausible situation, in which the discrete charged and neutral lipid mixtures move laterally in the membrane. Thus, we found the characteristic topological change caused by pore formation in a charged membrane (see Fig.~\ref{fig:morph_diagdam}). In addition, our MD simulation visualized details of the behavior at a molecular level. Figure~\ref{fig:detail} shows an example of the dynamics of lipid molecules around the pore-forming site within the charged-lipid-rich domain (red or dark) during the topological change. Interestingly, even before the onset of the pore formation at $t = 7.8\times7500\tau$, the orientation of several charged lipids became perpendicular to that of the surrounding lipids at $t = 7.4\times7500\tau$. The charged head group of the perpendicular lipid was embedded between the two leaflets of the bilayer membrane, probably because of the strong electrostatic repulsion among their charged head groups. This could trigger further instability in the membrane structure by interfering with the hydrophobic ordering attraction of the surrounding lipid chains, resulting in the pore formation. 

It should be noted that in the present calculation, the vesicle diameter was on the order of $10\,\mathrm{nm}$ due to the limitation imposed by computational cost. Previous studies have also performed similarly sized coarse-grained simulations to reproduce the dynamics of neutral lipid bilayer vesicles consisting of, for example, 4000--16000 lipids~\cite{Cooke}, 4096 lipids~\cite{Markvoort}, and 5072 lipids~\cite{Zheng}. The curvature effect, arising from the nature of the bilayer, in these typical coarse-grained calculations is larger than that of GUVs with negligible curvature, and the vesicles prefer to flatten, creating a pore. Because of the curvature effect, the values for GUVs can vary compared with those reported here. However, the magnitude of the bending energy per lipid molecule of $\sim 0.02k_\mathrm{B}T$, estimated from typical bilayer bending stiffness $\kappa\sim10k_\mathrm{B}T$ and curvature radius $\sim10\,\mathrm{nm}$, is much smaller than that of the attractive interaction potential of about $-20k_\mathrm{B}T$ and that of the electrostatic interaction potential of $\sim 0.8k_\mathrm{B}T$ for $n_{0}=100\,\mathrm{mM}$. Therefore, the deviation from the values for GUVs should be small, and the obtained values here should be relevant to GUVs and biomembranes on the microscale.

The consistency of the topological changes observed in this study with our previous observations of characteristic pore formation in charged lipid vesicles~\cite{Himeno2} validates the plausibility of our coarse-grained MD simulation of charged lipid vesicles. The critical salt concentration for the charge-induced pore formation in charged lipid vesicles was determined to be on the order of $100\,\mathrm{mM}$, which is the typical salt concentration to which biomembranes are exposed. In other words, this salt concentration level is necessary to maintain the closed membrane morphology. Considering that biomembranes contain only negatively charged lipids and thus are subjected to electrostatic repulsion, the topological and morphological stabilities of biomembranes could also be physically maintained by the screened electrostatic interactions at a salt concentration of $\sim 100\,\mathrm{mM}$. Moreover, the morphological dynamics of biomembranes, such as exo- and endocytosis, may be finely controlled by the precise tuning of the electrostatic repulsive forces via the salt concentration near the critical point. This should be investigated by future studies, including the effects of lipid asymmetry between the two leaflets, other charged proteins, explicit solvents, and so on, with comparisons to experiments and more detailed simulations.

\section{CONCLUSION}
\label{conclusion}

In this study, we focused on the dynamics of binary charged lipid vesicles, which are typical soft materials physiologically relevant to the basic structure of biomembranes. The important physicochemical phenomena in this system are the lateral phase separation and the morphological changes. By systematically changing the interaction parameters and salt concentration, we found that the lateral phase separation is significantly delayed and/or inhibited in the presence of electrostatic repulsion. The analysis of domain size in neutral and charged lipid vesicles suggested a different mechanism of phase separation in neutral and charged vesicles, where the domain-growth exponents for spinodal decomposition and nucleation growth were observed, respectively. Moreover, the longer term calculation unveiled the morphological dynamics, such as pore formation and further transformation into disk, string, and bicelle shapes, depending on the attractive interaction among lipids and salt concentration, as summarized in the phase diagram. The MD visualization suggested that the pore formation in the charged-lipid-rich domain was initiated by local disturbances of the orientational order of bilayer membranes, which appears prior to the pore formation. Our findings provide fundamental insights into the structure, dynamics, and stability of lipid bilayer membranes and vesicles containing charged lipids, which are broadly seen in biomembranes, as well as artificial membranes mimicking the functional features of biomembranes.


\begin{acknowledgments}
Most of the calculations were performed using the parallel computer ``SGI UV3000'' provided by Research Center for Advanced Computing Infrastructure at JAIST. H.I. acknowledges support from Grants-in-Aid for the Japan Society for the Promotion of Science (JSPS) Research Fellowship for Young Scientists (JP13J01297) and Sasakawa Scientific Research Grant from The Japan Science Society (28-233). N.S. acknowledges support from the Grant-in-Aid for Scientific Reserch on Innovative Areas ``{\it Molecular Robotics}" (JP15H00806) from the Ministry of Education, Culture, Sports, Science, and Technology of Japan (MEXT) and the Grant-in-Aid for Young Scientist (B) 
(JP26800222) from JSPS. 
\end{acknowledgments}

\newpage

\newpage
\bibliography{ref_revise}



\end{document}